# Temperature and thickness dependent magnetostatic properties of [Fe/Py]/FeMn/Py multilayers

D. M. Polishchuk[1,2], O. I. Nakonechna[1,3], Ya. M. Lytvynenko[1], V. Kuncser[4], Yu. O. Savina[5], V. O. Pashchenko[5], A. F. Kravets[1], A. I. Tovstolytkin[1,6*], V. Korenivski[2]

[1]Institute of Magnetism of the NAS of Ukraine and MES of Ukraine,
36-b Acad. Vernadskogo Blvd., Kyiv 03142, Ukraine

[2]Nanostructure Physics, Royal Institute of Technology, Stockholm 10691, Sweden

[3]Department of Physics, Taras Shevchenko National University of Kyiv,
64/13 Volodymyrska Str., Kyiv 01601, Ukraine

[4]National Institute of Materials Physics, Bucharest-Magurele 077125, Romania

[5]B. Verkin Institute for Low Temperature Physics and Engineering, NAS of Ukraine,
47 Nauky Avenue, Kharkiv 61103, Ukraine

[6]Faculty of Radiophysics, Electronics and Computer Systems, Taras Shevchenko National University of Kyiv, 64/13 Volodymyrska Str., Kyiv 01601, Ukraine

The magnetic properties of thin-film multilayers [Fe/Py]/FeMn/Py are investigated as a function of temperature and thickness of the antiferromagnetic FeMn spacer using SQUID magnetometry. The observed behavior differs substantially for the structures with 6-nm and 15-nm FeMn spacers. While the 15-nm-FeMn structure exhibits exchange pinning of both ferromagnetic layers in the entire measurement temperature interval from 5 to 300 K, the 6-nm-FeMn structure becomes exchange de-pinned in the vicinity of room temperature. The de-pinned state is characterized by a single hysteresis loop centered around zero field and having enhanced magnetic coercivity. The observed properties are explained in terms of finite-size effects and possibly ferromagnetic interlayer coupling through the thin antiferromagnetic spacer.

---

* Corresponding author: atov@imag.kiev.ua

# 1. Introduction

Nanostructured antiferromagnets (AFMs) have been the subject of increased attention for the use in spintronics owing to their rigidity to external magnetic fields, the absence of stray fields, and theoretically predicted strong spin transfer torque [1 – 3]. Traditionally thin-film AFMs have been used for creating the exchange bias in such spintronic nanostructured elements as spin valves and magnetic tunnel junctions. The development of new and improved functionalities of modern magnetoelectronic devices has advanced active research into nanostructured systems with improved exchange bias effect [1 – 9].

Exchange bias effect in a FM-AFM bilayer is commonly observed as a field offset of the hysteresis loop, which is also known as exchange pinning. This unidirectional exchange anisotropy can be created by cooling the bilayer in a constant magnetic field from a temperature that exceeds the Néel temperature ($T_N$) of AFM but is lower than the Curie temperature ($T_C$) of FM, to an operating temperature $T < T_N$ [2, 3]. As a result, the hysteresis loop exhibits a field offset that is opposite to the direction of the magnetic field applied during the cooling. Such offset of the hysteresis loop is usually characterized by the exchange bias field $H_b$. With increasing temperature, $H_b$ becomes zero at temperature $T = T_b < T_N$, also known as the blocking temperature [3]. However, the FM-AFM bilayer remains still exchange biased that is observed as the enhanced coercivity at $T < T_N$. The latter is another typical indication of an exchange-biased system.

In the first theoretical models [1, 10], the exchange bias was considered to be a purely surface phenomenon: while the magnetization of FM could be rotated under applied magnetic field, the spin structure of AFM was assumed to remain unperturbed [1]. However, the later studies have demonstrated that the spin structure of AFM can be affected that can subsequently lead to changes in the exchange bias [7–15]. In addition, during the rotation of the magnetization of the FM layer, a spiral spin structure can form in the AFM layer [16, 17]. One of the intriguing questions is how the FM layers interact through an AFM layer in a FM/AF/FM trilayer, which was addressed in a number of works [6,12,13,18–24].

In this work, the features of static magnetic properties caused by the interlayer coupling in FM1/AFM/FM2 structures with different AFM spacer thicknesses are studied. Our work, carried out on thin-film multilayers [Fe/Py]/FeMn/Py with the FeMn thickness $t = 6$ and 15 nm using detailed SQUID magnetometry at different temperatures, is aimed at filling in the gaps in understanding the static magnetic properties of FM1/AF/FM2 trilayers. Obtained results can be useful for engineering magnetic nanostructures for novel devices of antiferromagnetic spintronics.

## 2. Samples and Methods

FM1/AFM/FM2 trilayers used in our experiments had the composition of [Fe(6)/Py(3)]/FeMn($t$)/Py(5), where the layer thicknesses in "nm" are given in parenthesis, and FeMn and Py stand for $Fe_{50}Mn_{50}$ and $Ni_{80}Fe_{20}$ (permalloy) alloys. The Py(5) layer and the [Fe(6)/Py(3)] bilayer are the soft and hard ferromagnetic layers, respectively (hereafter Py and Fe*). The higher-coercive Fe is used to harden the soft Py material in Fe*. The Py(3) sublayer was needed to promote an fcc-texture at the [Fe/Py]-FeMn interface for the growth of the FeMn layer with desirable antiferromagnetic properties [25–27]. Since FeMn displays a strong thickness-dependence due to finite-size effects, we have fabricated trilayers with different thicknesses of the FeMn spacer ($t$ = 6 and 15 nm). Hereafter these samples will be referred as FeMn6 and FeMn15, respectively. The multilayers were deposited by DC magnetron sputtering (Orion, AJA Intern.) at room temperature. To induce a preferred magnetization direction, the samples were deposited and subsequently annealed at 250 °C in the presence of a saturating dc magnetic field applied in-plane.

The magnetization measurements as a function of temperature were performed using a SQUID magnetometer MPMS-XL5 Quantum Design, in the temperature range of 5–300 K, and in magnetic fields of up to 5 kOe applied in the film plane and parallel to the nominal axis of exchange bias.

## 3. Experimental results

Figure 1 shows typical magnetization $M$–$H$ loops of a trilayer Fe*/FeMn(15 nm)/Py (sample FeMn15) at selected temperatures. These curves are non-symmetrical with respect to both axes and each loop is a superposition of two minor hysteresis loops with distinctly different values of the exchange-bias field $H_b$ and the coercive field $H_c$. Since the Fe* layer has almost 3 times higher magnetic moment than the one of the Py layer, the minor loop that is taller in height and has smaller $H_b$ can be ascribed to Fe*. The other, strongly field-offset and shorter hysteresis loop belongs to the Py layer. Nonzero $H_b$ and increased $H_c$ both indicate that the two FM layers, Fe* and Py, are exchanged pinned in the entire temperature interval of the measurements. With increasing temperature, the total $M$–$H$ curve preserves the same double-loop shape, although the coercivity and exchange-bias field of both minor loops decrease considerably. The latter can be explained by the approaching of the blocking temperature of the FeMn(15 nm) spacer, which is expected to be higher than 300 K, as well as the Néel temperature. To note, the minor loop of Py at 10 K has a vertical brake, which results in a peculiar shape of the total $M$–$H$ loop: whereas the negative-field part of the minor loop of Py is observed at the bottom, the positive-field part of the loop is at the top. This is because $H_c$ of Py is substantially larger than that of Fe*, and importantly $H_c > H_b$ for the Py layer.

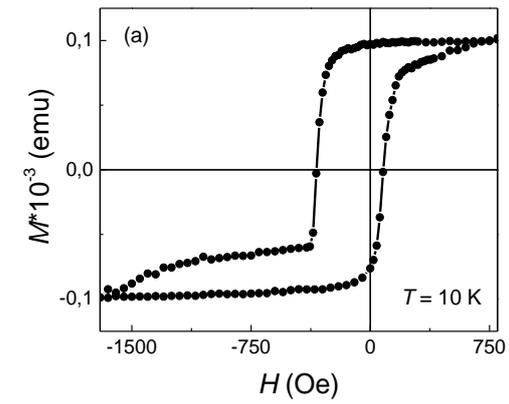

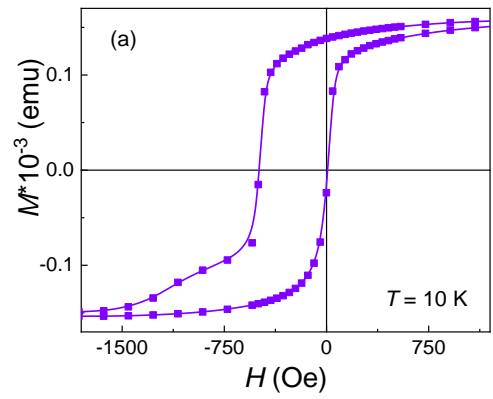

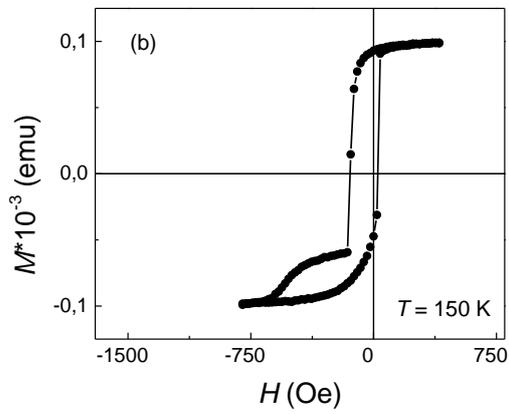

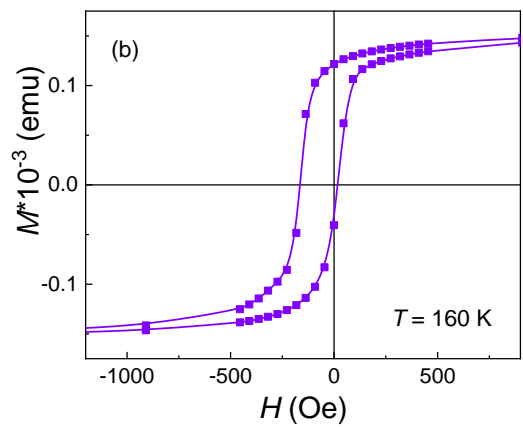

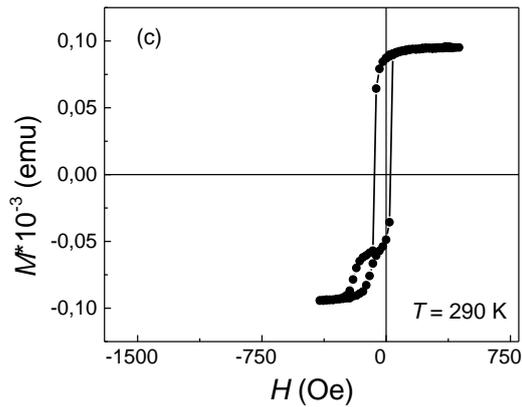

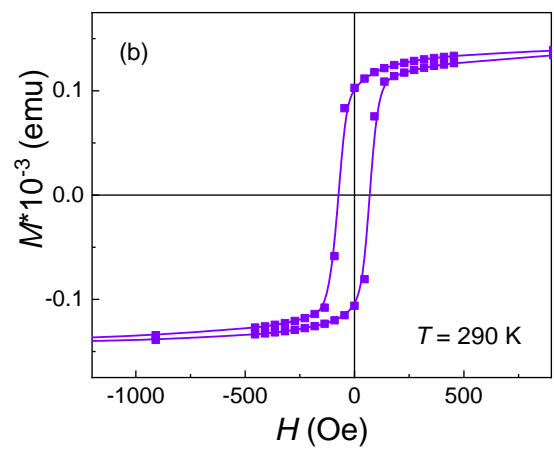

**Fig. 1.** *M–H* curves for Fe*/FeMn(15 nm)/Py (sample FeMn15) obtained at different temperatures.

**Fig. 2.** *M–H* curves for Fe*/FeMn(6 nm)/Py (sample FeMn6) obtained at different temperatures.

Figure 2 demonstrates $M$–$H$ loops for a trilayer Fe*/FeMn(6 nm)/Py (sample FeMn6) at select temperatures. In contrast to the FeMn15 sample, a double loop of FeMn6 at 10 K transforms into a single loop at $T \geq 160$ K. Importantly, this single loop reveal non-zero exchange bias field, $H_b \approx 25$ G at $T = 160$ K. A single hysteresis loop in such trilayers, where Fe* and Py have intrinsically different coercive fields, is the indication of a relatively strong interlayer coupling between the outer FM layers, Fe* and Py, when the structure behaves like a single FM film. This coupling is mediated through the 6-nm FeMn layer that has quite strong antiferromagnetic order at lower temperatures indicated by non-zero $H_b$. A single loop is also observed at room temperature but now it loses its exchange offset, $H_b = 0$. This implies that the Fe* and Py layers remain coupled while the antiferromagnetic order in the FeMn spacer becomes considerably weaker. It is noteworthy that 6 nm is the critical thickness, when thin-film FeMn loses its strong anisotropic and AFM properties at room temperature as was shown in a series of reports [28 – 30].

Figure 3 shows a comparative analysis of hysteresis loops for the FeMn6 and FeMn15 samples at 10 K and 290 K. In contrast to FeMn6, the FeMn15 structure has much stronger exchange-bias offset of the minor loops of both Fe* and Py, which is also preserved at room temperature. This can be explained by much stronger AFM order in the 15-nm FeMn spacer. The weaker antiferromagnetism of the 6-nm FeMn spacer can be ascribed to the finite-size effect, when the thickness is not sufficiently large for building strong long-range AFM order.

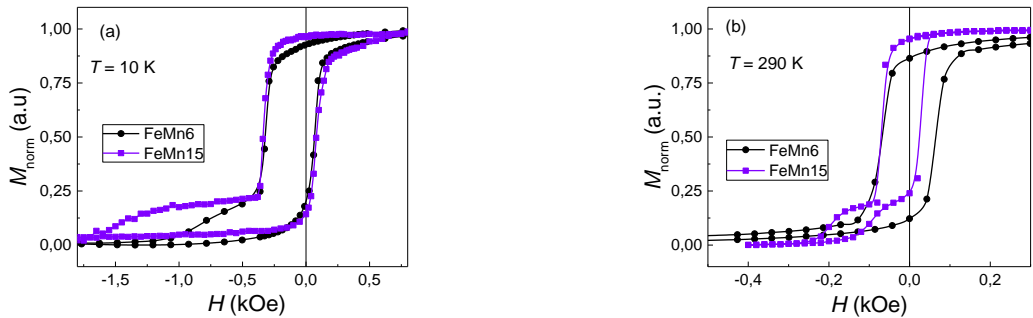

**Fig. 3.** Normalized hysteresis loops obtained at 10 K (*a*) and 290 K (*b*).

Due to the complex shape of the $M$–$H$ loops, the values of $H_c$ and $H_b$ were determined using derivatives $dM/dH$, as shown in Fig. 4 for the FeMn15 structure at 290 K. On the basis of the center fields of the peaks in $dM/dH(H)$ curves, one can calculate the coercive field and exchange bias field of the Fe* ($H_{c1}$, $H_{b1}$) and Py ($H_{c2}$, $H_{b2}$) minor loops. The results of such data processing are shown in Fig. 5.

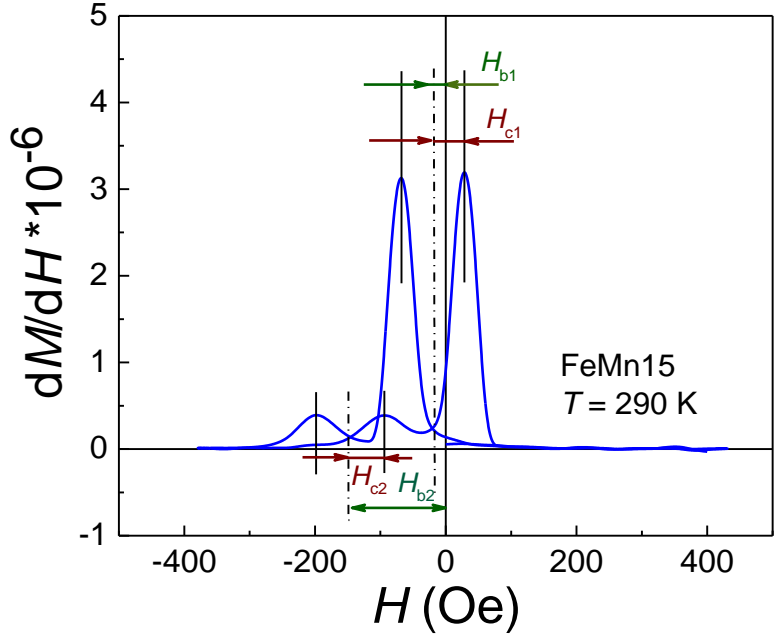

**Fig. 4.** First derivate d$M$/d$H$ for the FeMn15 sample at 290 K. $H_{b1,c1}$ and $H_{b2,c2}$ indicate exchange bias/coercive fields for the Fe* and Py layers, respectively.

The temperature dependences of the exchange bias $H_{b1}$ and $H_{b2}$ exhibit pronounced differences for the FeMn6 and FeMn15 structures (Fig. 5, *a*). The most vivid difference is in $H_{b2}$ of the Py layer. $H_{b2}$ is non-zero in the entire temperature interval for FeMn15, and is much larger than that for FeMn6 at lower temperatures. Since the Py layer is rather thin and soft FM, it is very sensitive to the exchange bias effect. This explains such a big difference in the $H_{b2}$ as due to much weaker exchange bias effect and subsequently weaker antiferromagnetism in the 6-nm FeMn layer. Remarkably, for the FeMn6 structure, $H_{b2}$ abruptly drops down at ~60 K that manifests the onset of the coupling between the outer Py and Fe* layers at $T > 60$ K. Interestingly, the values of $H_{b1}$ are quite close for the two structures at low temperatures, but diverge considerably with increasing temperature that can be related to different properties of the 6-nm and 15-nm FeMn spacers. Less pronounced temperature dependence for $H_{b1}$ can be explained by much larger magnetic moment of Fe* and hence lower sensitivity to the exchange bias effect.

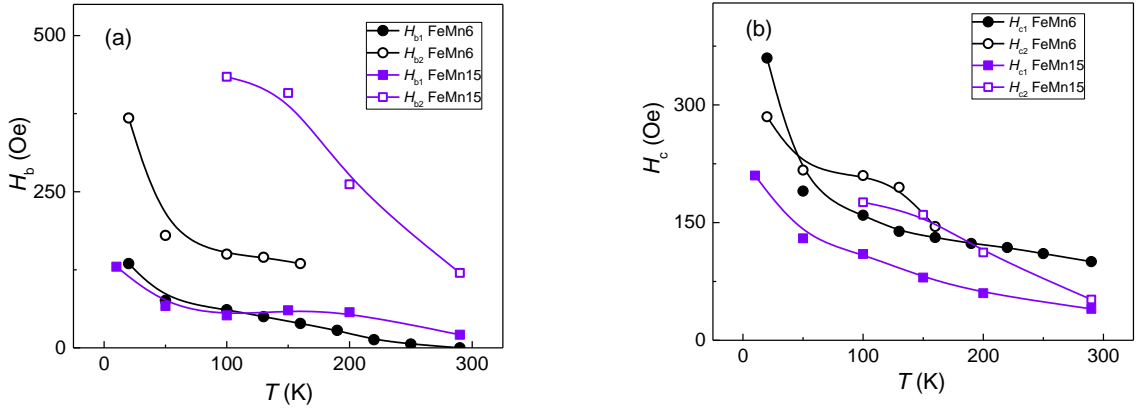

**Fig. 5.** Temperature dependences of $H_b$ (*a*) and $H_c$ (*b*).

The multilayers were purposely designed to have ferromagnetic layers with different intrinsic coercivities – hard Fe* = [Fe/Py] ($H_{c1}$) and soft Py ($H_{c2}$). Intrinsically, $H_{c1} \sim 100$ Oe $> H_{c2} \sim 5$ Oe. However, for the actual Fe*/FeMn/Py trilayers, $H_{c1} < H_{c2}$, which can be explained by very strong exchange bias effect on the thinner and softer Py layer from the FeMn spacer. As seen in Fig. 5, *b*, the inequality $H_{c1} < H_{c2}$ holds in the whole temperature interval for FeMn15 and at low temperatures for FeMn6. At higher temperatures, the single *M–H* loop for FeMn6 indicates a ferromagnetic coupling between the outer ferromagnetic layers, when the magnetization switching of the Fe* layer at $H_{c1}$ drives the switching of the Py layer.

**Conclusions**

For the FeMn6 sample, a single-loop hysteresis is observed at around room temperature. In this temperature range, the exchange-bias field is relatively small ($H_b \sim 0$) but the coercivity is enhanced and exceeds that of the thick-spacer sample (FeMn15). This behavior is consistent with the spacer preserving partial AFM ordering while losing its magnetic anisotropy, whose value is not sufficient to exchange pin the ferromagnets at this elevated temperature.

As the temperature is decreased, the single hysteresis loop develops a field offset, which indicates enhanced anisotropy and, as a result, stiffening of the exchange pinning. At a certain temperature, the single loop splits into two minor loops and the behavior transitions toward individual pinning of the Fe* and Py layers, with insignificant interlayer FM coupling.

For thick AFM spacers (FeMn15 sample), the trilayer demonstrates relatively strong exchange pinning and negligible interlayer coupling, manifesting a strong exchange-bias field and enhanced coercivity for both ferromagnetic layers at low temperatures.

.

**Acknowledgements**

Support from the National Academy of Sciences of Ukraine (Projects No. 0119U100469 and 0120U100457), the Swedish Research Council (VR:2018-03526) and Olle Engkvist Foundation (2020:207-0460) are gratefully acknowledged.

**References**

1. W.H. Meiklejohn, *J. Appl. Phys.* 33, 1328 (1962). https://doi.org/10.1063/1.1728716
2. J. Nogués, Ivan K. Schuller, *J. Magn. Magn Mater.* 192, 203 (1999) https://doi.org/10.1016/S0304-8853(98)00266-2.
3. K. O'Grady, L.E. Fernandez-Outon, G. Vallejo-Fernandez, *J. Magn. Magn Mater.* 322, 883 (2010) https://doi.org/10.1016/j.jmmm.2009.12.011
4. S. Parkin, X. Jiang, C. Kaiser, A. Panchula, K. Roche, M. Samant, *Proc. IEEE* 91, 661 (2003) https://doi.org/10.1109/JPROC.2003.811807
5. S.D. Bader, S.S.P. Parkin, *Annu. Rev. Condens. Matter Phys.* 7, 71 (2010) https://doi.org/10.1146/annurev-conmatphys-070909-104123
6. P. Miltényi, M. Gierlings, J. Keller, B. Beschoten, G. Güntherodt, U. Nowak, K. D. Usadel, *Phys. Rev. Lett.* 84, 4224 (2000) https://doi.org/10.1103/PhysRevLett.84.4224
7. G. Malinowski, M. Hehn, S. Robert, O. Lenoble, A. Schuhl, P. Panissod, *Phys. Rev. B* 68, 184404 (2003) https://doi.org/10.1103/PhysRevB.68.184404
8. J.D. Dutson, C. Huerrich, G. Vallejo-Fernandez, L.E. Fernandez-Outon, G. Yi, S. Mao et al., *J. Phys. D Appl. Phys.* 40, 1293 (2007) https://doi.org/10.1088/0022-3727/40/5/S16 .
9. X.Z. Zhan, G. Li, J.W. Cai, T. Zhu, J.F.K. Cooper, C.J. Kinane, S. Langridge, *Sci. Rep.* 9, 6708 (2019) https://doi.org/10.1038/s41598-019-43251-1
10. M. L. Pankratova, A. S. Kovalev; *Low Temp. Phys*. 41, 838 (2015) https://doi.org/10.1063/1.4934546
11. W.H. Meiklejohn, C.P. Bean, *Phys. Rev.* 102, 1413 (1956) https://doi.org/10.1103/PhysRev.105.904
12. R. Morales, Z.P. Li, J. Olamit, K. Liu, J.M. Alameda, I.K. Schuller, *Phys. Rev. Lett*. 102, 097201 (2009), https://doi.org/10.1103/PhysRevLett.102.097201
13. A.V. Svalov, G.V. Kurlyandskaya, V.N. Lepalovskij, P.A. Savin, V.O. Vas'kovskiy, Superlattice. Microst. 83, 216 (2015) https://doi.org/10.1016/j.spmi.2015.03.040.
14. X.Z. Zhan, G. Li, J.W. Cai, T. Zhu, J.F.K. Cooper, C.J. Kinane, S. Langridge, *Sci. Rep*. 9, 6708 (2019), https://doi.org/10.1038/s41598-019-43251-1.
15. P.D. Kulkarnia, P.V. Sreevidya, J. Khan, P. Predeep, H.C. Barshilia, P. Chowdhury, *J. Magn. Magn Mater*. 472, 111 (2019) https://doi.org/10.1016/j.jmmm.2018.10.026.

16. D. Mauri, H.C. Siegmann, P.S. Bagus, E. Kay, J. Appl. Phys. 62, 3047 (1987) https://doi.org/10.1063/1.339367.
17. M.D. Stiles, R.D. McMichael, *Phys. Rev. B* 59, 3722 (1999) https://doi.org/10.1103/PhysRevB.59.3722.
18. A. F. Kravets, O. V. Gomonay, D. M. Polishchuk, Yu. O. Tykhonenko-Polishchuk, T. I. Polek, A. I. Tovstolytkin, V. Korenivski, *AIP Advances* 7, 056312 (2017) https://doi.org/10.1063/1.4975694.
19. Y. Xu, Q. Ma, J.W. Cai, L. Sun, *Phys. Rev. B* 84 (2011), 054453 https://doi.org/10.1103/PhysRevB.84.054453.
20. F.Y. Yang, C.L. Chien, *Phys. Rev. Lett.* 85, 2597 (2000) https://doi.org/10.1103/PhysRevLett.85.2597.
21. C.W. Leung, M.G. Blamire, *J. Appl. Phys*. 94 (2003) 7373 https://doi.org/10.1063/1.1624479.
22. C.W. Leung, M.G. Blamire, *Phys. Rev. B* 72, 054429 (2005), https://doi.org/10.1103/PhysRevB.72.054429.
23. D.N.H. Nam, W. Chen, K.G. West, D.M. Kirkwood, J. Lu, S.A. Wolf, *Appl. Phys. Lett.* 93, 152504 (2008), https://doi.org/10.1063/1.2999626.
24. M.Y. Khan, C.-B. Wu, W. Kuch, Phys. Rev. B 89 (2014), 094427 https://doi.org/10.1103/PhysRevB.89.094427.
25. K. Nishioka, J. *Appl. Phys.* 80, 4528 (1996) https://doi.org/10.1063/1.363433.
26. K.-Y. Kim, H.-C. Choi, S.-Y. Jo, C.-Y. You, *J. Appl. Phys*. 114, 073908 (2013) https://doi.org/10.1063/1.4818955.
27. W.J. Antel, F. Perjeru, G.R. Harp, *Phys. Rev. Letters* 83 (1999) 1439 https://doi.org/10.1103/PhysRevLett.83.1439.
28. K. Lenz, S. Zander, W. Kuch, *Phys. Rev. Letters* 98 (2007) 237201 https://doi.org/10.1103/PhysRevLett.98.237201
29. P. Merodio, A. Ghosh, C. Lemonias, E. Gautier, U. Ebels, M. Chshiev, et al., *Appl. Phys. Letters* 104 (2014) 032406 https://doi.org/10.1063/1.4862971
30. D.M. Polishchuk, T.I. Polek, V.Yu. Borynskyi, A.F. Kravets, A.I. Tovstolytkin, A.M. Pogorily, V. Korenivski, *Low Temperature Physics* 46 (2020) 813 https://doi.org/10.1063/10.0001547